\documentclass[aps,prd,notitlepage,showpacs,nofootinbib,superscriptaddress]{revtex4-1}

\pdfoutput=1
\usepackage[usenames,dvipsnames]{xcolor}  
\usepackage{colortbl} 
\usepackage{graphicx} 
\usepackage[compat=1.1.0]{tikz-feynman}
\definecolor{lightblue}{rgb}{0.93, 0.95, 1.0} 
\definecolor{lightgray}{gray}{0.9} 
\usepackage{multirow}
\usepackage{bm}
\usepackage{bbold}
\usepackage{adjustbox}

\usepackage{latexsym}
\usepackage{amsthm}
\usepackage{amsmath}
\usepackage{slashed}
\usepackage{xfrac}
\usepackage{wrapfig}
\usepackage{amssymb}
\usepackage{cancel}
\usepackage{multirow}
\usepackage{pst-3d}                  
\usepackage{pst-grad}                
\usepackage{pst-node}                
\usepackage{pst-slpe}                
\usepackage{pst-plot}                %
\usepackage{pst-coil}                %
\usepackage{pst-math}                %
\usepackage{pstricks-add}            %
\usepackage{rotating}
\usepackage{geometry}
\usepackage{hhline}
\usepackage{amsmath}
\usepackage{float}
\usepackage{fancybox}
\usepackage[utf8]{inputenc}

\usepackage[colorlinks=true,citecolor=darkred,urlcolor=darkred, pdfborder={0 0 0}]{hyperref}
\usepackage[normalem]{ulem}

\usepackage{color,soul}
\setul{0.5ex}{0.3ex}
\setulcolor{blue}

\usepackage{braket}
\definecolor{linkcolor}{rgb}{0,0,0.5}
\allowdisplaybreaks
\allowdisplaybreaks[1]
\allowdisplaybreaks[2]

\definecolor{greenLinks}{rgb}{0, 0.6, 0}
\definecolor{blueLinks}{rgb}{0, 0, 0.6}
\definecolor{redLinks}{rgb}{0.6, 0, 0}
\definecolor{tempText}{rgb}{0.55, 0.10,0.67}
\definecolor{eprintLinks}{rgb}{0.4, 0.4, 0.4}
\definecolor{journalLinks}{rgb}{0.6, 0, 0}
\definecolor{mygreen}{rgb}{0.0, 0.5, 0.1}
\newcommand {\ignore}[1]{}

\usepackage{soul}
\definecolor{mightnightblue}{RGB}{25,25,112}

\definecolor{brown}{rgb}{0.59, 0.29, 0.0}

\definecolor{darkred}{rgb}{0.6,0,0}

\def\lsim{\mathrel{\rlap{\lower4pt\hbox{\hskip1pt$\sim$}}
    \raise1pt\hbox{$<$}}}
\def\gsim{\mathrel{\rlap{\lower4pt\hbox{\hskip1pt$\sim$}}
    \raise1pt\hbox{$>$}}}

\def\U1s{$\mathrm{U_{1}^{(a)}\otimes U_{1}^{(b)}\otimes U_{1}^{(c)}\otimes U_{1}^{(d)}\otimes U_{1}^{(e)}}$ }
\def\3211{$\mathrm{SU(3) \times SU(2)_L \times U(1)_R \times U(1)_{B-L}}$ }
\def\321{$\mathrm{SU(3) \times SU(2) \times U(1)}$ }
\def\422{$\mathrm{SU(4) \times SU(2) \times SU(2)_R}$ }

\newcommand{\AddrAHEP}{%
  AHEP Group, Institut de F\'{i}sica Corpuscular --
  CSIC/Universitat de Val\`{e}ncia, Parc Cient\'ific de Paterna.\\
 C/ Catedr\'atico Jos\'e Beltr\'an, 2 E-46980 Paterna (Valencia), Spain}
 
 \newcommand{\AddrMPI}{Max-Planck-Institut f\"ur Kernphysik, Saupfercheckweg 1, 69117 Heidelberg, Germany\vspace{0.2cm}}
 
  \newcommand{\AddrIPN}{Departamento de F\'isica, Centro de Investigaci\'on y de Estudios Avanzados del IPN Apdo. Postal 14-740 07000 Ciudad de M\'exico, M\'exico\vspace{0.2cm}}

\def\black{\color{black}{}}

\newcommand{\jv}[1]{{\leavevmode\color{magenta}{#1}}}

\begin{document}
\title{\boldmath \color{BrickRed} Leptonic neutral-current probes in a short-distance DUNE-like setup}

\author{Salvador Centelles Chuli\'{a}}\email{chulia@mpi-hd.mpg.de}
\affiliation{\AddrMPI}
\author{O. G. Miranda}\email{omar.miranda@cinvestav.mx}
\affiliation{\AddrIPN}
\author{Jose W.F. Valle}\email{valle@ific.uv.es}
\affiliation{\AddrAHEP}

\begin{abstract}
\noindent

Precision measurements of neutrino-electron scattering may provide a viable way to test the non-minimal form of the charged and neutral current
weak interactions within a hypothetical near-detector setup for the Deep Underground Neutrino Experiment (DUNE).
Although low-statistics, these processes are clean and provide information complementing the results derived from oscillation studies.
They could shed light on the scale of neutrino mass generation in low-scale seesaw schemes.
\end{abstract}

\maketitle

\section{Introduction}
\label{sec:intro}

Ever since the historic discovery of neutrino oscillations~\cite{Kajita:2016cak,McDonald:2016ixn} indicating the need for neutrino masses, most  progress in neutrino physics has relied on experiments involving charged current (CC) processes.
However, the observation of coherent elastic neutrino-nucleus scattering (CEvNS)~\cite{COHERENT:2017ipa,zenodo} proposed long ago in the pioneer papers~\cite{Freedman:1973yd, Drukier:1984vhf}, has prompted a neutral current (NC) \textit{revival} and the recognition that the NC can provide an interesting and complementary way to study neutrinos.

Moreover, it has long been known from theory that, if neutrino mass generation is mediated by heavy neutrino exchange, the structure of  both the CC and NC is non-trivial~\cite{Schechter:1980gr,Schechter:1981cv}.
Indeed, the mixing matrix $K$ characterizing the leptonic CC weak interaction that describes oscillations is not unitary, while the NC interaction of mass-eigenstate neutrinos involves a matrix $P$ that deviates from the unit matrix. 
Moreover, these two matrices are related~\cite{Schechter:1980gr,Schechter:1981cv}. \par 
Although the effect of non-unitary neutrino mixing was first discussed in the context of  astrophysical neutrino propagation~\cite{Valle:1987gv,Nunokawa:1996tg,Grasso:1998tt},
it can be phenomenologically relevant in earth-bound experiments. 
This happens in the context of genuine low-scale seesaw schemes, such as the inverse~\cite{Mohapatra:1986bd,GonzalezGarcia:1988rw} or the linear seesaw mechanism~\cite{Akhmedov:1995ip,Akhmedov:1995vm,Malinsky:2005bi}, 
leading to potentially sizeable deviations from the conventional leptonic weak currents with unitary CC mixing.
These corrections are expressed as power series in the parameter $\varepsilon = \mathcal{O}(Y v / M)$, where $M$ is the mediator mass scale and $v$ is the SM vacuum expectation value (vev). Although small, we stress that $\varepsilon$ can be non-negligible within low-scale realisations of the seesaw.
The effects of unitarity violation and the new associated neutrino phenomena have been extensively explored in the recent literature, mainly devoted to charged-current processes on hadronic probes~\cite{Escrihuela:2015wra,Escrihuela:2016ube,Fong:2016yyh,Ge:2016xya,Miranda:2016ptb,Miranda:2016wdr,Fong:2017gke,Miranda:2019ynh,Miranda:2020syh,Martinez-Soler:2021sir,Rahaman:2021cgc,Soumya:2021dmy,Kaur:2021rau,Wang:2021rsi,Chatterjee:2021xyu,Gariazzo:2022evs,Aloni:2022ebm,Sahoo:2023mpj,Celestino-Ramirez:2023zox,Miranda:2021kre,Schwetz:2021thj,Schwetz:2021cuj,Tang:2021lyn,Arrington:2022pon,Capozzi:2023ltl,Soleti:2023hlr}. \\

Here we explore the potential of leptonic probes such as the scattering process $\nu + e^- \to \nu + e^-$ as a potentially viable way to test the non-minimal form of the CC and NC weak interactions. 
This has already been discussed in~\cite{Miranda:2018yym, Coloma:2021uhq} in the context of hadronic probes.
In this paper we speculate that leptonic probes, too, may be useful within a DUNE-like near-detector setup.
Although low-statistics, these processes are clean and provide information complementing the results derived from oscillation studies~\cite{deSalas:2020pgw},
shedding light on the scale of neutrino mass generation within low-scale seesaw schemes.

\section{Theory preliminaries} 
\label{prel}
The most general CC weak interaction of massive neutrinos is described by a rectangular matrix $K$~\cite{Schechter:1980gr}. A short summary of the main features and notation is as follows. 
In the basis where the charged lepton mass matrix is diagonal, the upper blocks are simply the first 3 rows of the neutrino mixing matrix~\cite{Schechter:1981cv}. We can also define the relevant sub-block as
\begin{equation}
    K = \left(\begin{matrix} N & S \end{matrix}\right) \\
\end{equation}
With $3$ active neutrino flavours, $N$ is a $3\times 3$ matrix, while $S$ is a $ 3\times m$ matrix, with $m$ the number of fermionic singlets that mix with the active neutrinos~\footnote{In the most general SM-seesaw one can have any number of ``right-handed'' neutrino mediators, since they are gauge singlets~\cite{Schechter:1980gr}.}. The small block $S \sim \mathcal{O}\left(\varepsilon\right)$ is the seesaw expansion matrix~\cite{Schechter:1981cv}. 
Notice that $K K^\dagger = I_{3\times 3}$ and therefore $N N^\dagger = 1 - S S^\dagger \sim 1- \mathcal{O}\left(\varepsilon^2\right)$. On the other hand, the matrix $P$ describing the NC-neutrino interactions~\cite{Schechter:1980gr} is given by
\begin{equation}
    P = K^\dagger K \neq I~.
\end{equation}
If the energy of a given process is much lower than the masses of the heavy mediators these will not be produced. 
For example, the heavy states will not take part in oscillation experiments. 
Then, effectively, only the first $3\times 3$ blocks of $K$ and $P$ will play a role in the weak interactions, i.e. $N$ in the charged current and $N^\dagger N$ in the neutral current.  
This would signal the presence of unitarity violation in the neutrino mixing matrix~\cite{Valle:1987gv}
whose general description is given in~\cite{Escrihuela:2015wra}. 
A systematic approach to the (non-unitary) matrix $N$ can be derived from the seesaw expansion~\cite{Schechter:1981cv}, as the lower triangular parametrization proposed in~\cite{Escrihuela:2015wra}~\footnote{An alternative description and its relationship with Eq.~\ref{eq:Nparam} is discussed in Ref.~\cite{Blennow:2016jkn}}.
\begin{eqnarray}
\label{eq:Nparam}
  N &=& \left(\begin{matrix}
      \alpha_{11} &     0       & 0 \\
      \alpha_{21} & \alpha_{22} & 0 \\
      \alpha_{31} & \alpha_{31} & \alpha_{33}
  \end{matrix}\right) \cdot U 
\end{eqnarray}
 Besides the $3\times 3$ unitary matrix $U$ used to describe neutrino mixing in the conventional unitary case, one has the triangular pre-factor characterized by 3 diagonal $\alpha_{ii}$ ($i=1,2,3$), real and close to 1, and 3 non-diagonal $\alpha_{ij}$ ($i\neq j$) which are small but complex.
This is a convenient and complete description of non-unitarity.
By construction, $N$ and $S$ must satisfy the relation $N N^\dagger = 1 - S S^\dagger$, hence $N N^\dagger \sim 1 - \mathcal{O}\left(\varepsilon^2\right)$. Explicitly,
\begin{equation}
    N N^\dagger  = \begin{pmatrix}
        \alpha_{11}^2           & \alpha_{11} \alpha_{21}^*         & \alpha_{11} \alpha_{31}^* \\
        \alpha_{11} \alpha_{21} & \alpha_{22}^2 + |\alpha_{21}|^2   & \alpha_{22} \alpha_{32}^* + \alpha_{21} \alpha_{31}^* \\
        \alpha_{11} \alpha_{31} & \alpha_{22} \alpha_{32} + \alpha_{21}^* \alpha_{31}          & \alpha_{33}^2 + |\alpha_{31}|^2 + |\alpha_{32}|^2
    \end{pmatrix}~,
\end{equation}
from where one can read off the strength of the $\alpha_{ij}$ in terms of the small seesaw expansion parameter $\varepsilon$.  %
Indeed, within the seesaw paradigm, the leading deviation from the standard form of the CC mixing matrix lies in the diagonal entries $N N^\dagger \sim 1 - \mathcal{O}\left(\varepsilon^2\right)$ implying that, for example, $\alpha^2_{11}\sim 1-\mathcal{O}\left(\varepsilon^2\right)$. Hence $\alpha_{11}\alpha_{21}\sim \mathcal{O}\left(\varepsilon^2\right)$, so that $\alpha_{21}\sim\mathcal{O}\left(\varepsilon^2\right)$ and
\begin{eqnarray}
   \label{eq:epsorder1} \alpha_{ii}^2 &\sim& 1-\mathcal{O}\left(\varepsilon^2\right) \\
    \label{eq:epsorder2} |\alpha_{ij}|^2 &\sim& \mathcal{O}\left(\varepsilon^4\right) , \hspace{0.5cm} i\neq j~.
\end{eqnarray}
One sees that the strength of the off-diagonal $\alpha$'s is suppressed relative to the deviations of the flavor-diagonal ones from their SM values. 
In other words, in the seesaw expansion the $0^\text{th}$ order corresponds to the unitary limit, 
the $1^\text{st}$ order gives only diagonal flavour-conserving effects, while
the genuine flavour-violating effects non-unitary corrections only come at $2^\text{nd}$ order.
Notice also that this behavior is consistent with the validity of the well-known triangle inequality $|\alpha_{ij}| \leq \sqrt{(1-\alpha^2_{ii})(1-\alpha^2_{jj})}$ \cite{Escrihuela:2015wra}.
\par

Additionally, it is important to notice that unitarity violation leads to a redefinition of the Fermi constant, extracted from the $\mu^-$ lifetime assuming the SM.
In the presence of non-unitarity the measured quantity would be the effective muon decay coupling $G_\mu$. Since the W-boson vertices are modified by the non-unitarity parameters one finds
\begin{align}
G_\mu &= 1.1663787(6) \times 10^{-5} \text{ GeV}^{-2} & \text{(effective $\mu^-$ decay constant, \cite{Workman:2022ynf})} \\
G^2_\mu &= G^2_F (N N^\dagger)_{ee} (N N^\dagger)_{\mu \mu}
\end{align}
And therefore
\begin{equation}
\label{eq:GF}
    1 \leq \frac{G_F^2}{G_\mu^2} = \frac{1}{(N N^\dagger)_{ee} (N N^\dagger)_{\mu \mu}} \approx 3 - \alpha_{11}^2 - \alpha_{22}^2 \sim 1+ \mathcal{O}\left(\varepsilon^2\right)
\end{equation}
Therefore, in the presence of non-unitarity any process proportional to $G_F^2$ will get an ``enhancement''. 
This is counter-intuitive, because naively one expects less events than in the SM if the mixing is non-unitary, due to kinematically inaccessible heavy states.  
The reduction of the event number due to non-unitarity and the ``increase'' due to the redefinition of $G_F$ compete with each other, 
so that in some cases one can achieve $\mathcal{N}_\text{NU}/\mathcal{N}_\text{U} = 1$ even in the presence of non-unitarity. \par 

\section{$\nu_\mu - e^-$ scattering in the presence of non-unitarity at zero distance} 
\label{sec:scatt}
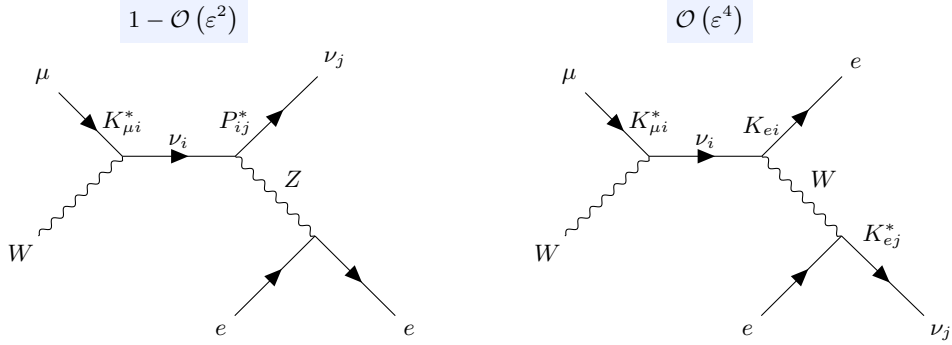
\begin{figure}[t!]
\begin{tikzpicture}\hspace{-3cm}
\begin{feynman}
    \vertex (a) {\(\mu\)};
    \vertex [below right=of a] (b) ;
    \vertex [right=of b] (c) ;
    \vertex [above right=of c] (d) {\(\nu_j\)};
    \vertex [below left=of b] (e) {\(W\)};
    \vertex [below right=of c] (f);
    \vertex [below left=of f] (g) {\(e\)};
    \vertex [below right=of f] (h) {\(e\)};
    \diagram* {
      (a) -- [fermion] (b) -- [fermion, edge label=\(\nu_i\)] (c) -- [fermion] (d),
      (e) -- [boson] (b),
      (c) -- [boson, edge label=\(Z\)] (f),
      (g) -- [fermion] (f) -- [fermion] (h),
    };
    \vertex [above=0.5em of b] {\(K_{\mu i}^*\)};
    \vertex [above=0.5em of c] {\(P_{ij}^*\)};
    \node [above=of b, yshift=0em, xshift=2.5em,fill=lightblue]{$1-\mathcal{O}\left(\varepsilon^2\right)$};  
\end{feynman} 
 \begin{feynman}\hspace{7cm}
    \vertex (a) {\(\mu\)};
    \vertex [below right=of a] (b) ;
    \vertex [right=of b] (c) ;
    \vertex [above right=of c] (d) {\(e\)};
    \vertex [below left=of b] (e) {\(W\)};
    \vertex [below right=of c] (f);
    \vertex [below left=of f] (g) {\(e\)};
    \vertex [below right=of f] (h) {\(\nu_j\)};
    \diagram* {
      (a) -- [fermion] (b) -- [fermion, edge label=\(\nu_i\)] (c) -- [fermion] (d),
      (e) -- [boson] (b),
      (c) -- [boson, edge label=\(W\)] (f),
      (g) -- [fermion] (f) -- [fermion] (h),
    };
    \vertex [above=0.5em of b] {\(K_{\mu i}^*\)};
    \vertex [above=0.5em of c] {\(K_{e i}\)};
    \vertex [right=0.5em of f] {\(K^*_{ej}\)};
\node [above=of b, yshift=0em, xshift=2.5em, fill=lightblue]{$\mathcal{O}\left(\varepsilon^4\right)$};
  \end{feynman}
  \end{tikzpicture}
\caption{Feynman diagrams for \(\nu_\mu + e^-\)  scattering, where the $\nu_\mu$ is produced via the usual CC vertex, while the final-state detection involves the NC (left diagram), with a sub-leading CC contribution, of order \(\varepsilon^4\) (right diagram).}
\label{fig:numudiags} 
\end{figure}

We now turn our attention to the scattering of a muon neutrino on an electron target in the presence of non-unitarity. The relevant Feynman diagrams, given in Figs.~\ref{fig:numudiags} and \ref{fig:CCdiags1}, describe the elastic electron-neutrino  scattering and the neutrino-induced muon production processes, respectively.

\subsection{Neutrino-electron elastic scattering} 
\label{subsec3a}
At tree level the two relevant diagrams are given in Fig.~\ref{fig:numudiags}. 
The vertex on the left of each diagram corresponds to CC $\nu_\mu$ production,  
while the vertex on the right depicts either the NC and CC detection modes, respectively. 
Within the three neutrino paradigm, with unitary lepton mixing, the CC channel (right panel of Fig.~\ref{fig:numudiags}) 
does not contribute at zero distance. 
In such case, the process is pure NC (left pannel of Fig.~\ref{fig:numudiags}) and the differential cross-section is given by 
\begin{eqnarray}
\label{eq:diffcrossnumu}
    \left(\frac{d\sigma}{dT}\right)^\text{SM} = \frac{2 G_\mu^2 m_e}{\pi} \left(g_L^2 + g_R^2 \left(1- \frac{T}{E\nu}\right)^2 - g_L g_R  \frac{m_e T}{E_\nu^2}\right)
\end{eqnarray}
where $g_L = -1/2 + \sin^2{\theta_\text{W}}$ and $g_R = \sin^2{\theta_\text{W}}$.
However, in the presence of non-unitarity the CC contribution is in general non-zero. In this case, the cross section is replaced by 
\begin{equation}
\label{eq:diffcrossnumuNU}
     \colorbox{lightblue}{$\displaystyle\left(\frac{d\sigma}{dT}\right)^\text{NU} =  \frac{\mathcal{P}_{\mu e}^{\text{NC}} }{(N N^\dagger)_{ee} (N N^\dagger)_{\mu \mu}} \left(\frac{d\sigma}{dT}\right)^\text{SM} + \frac{2 m_e G_\mu^2}{\pi} \frac{\mathcal{R}e\left[\mathcal{P}_{\mu e}^\text{int}\right]}{(N N^\dagger)_{ee} (N N^\dagger)_{\mu \mu}}\left\{\frac{\mathcal{P}_{\mu e}^{\text{CC}}}{\mathcal{R}e\left[\mathcal{P}_{\mu e}^\text{int}\right]}  + 2 g_L -  g_R \frac{m_e T}{E_\nu^2}\right\}$} ,
\end{equation}
where the probability factors are given by
\begin{eqnarray}
\mathcal{P}_{\mu e}^{\text{NC}} = (N N^\dagger N N^\dagger N N^\dagger)_{\mu \mu} \\
\mathcal{P}_{\mu e}^{\text{CC}} = (N N^\dagger)_{\mu e} (N N^\dagger)_{e\mu} (N N^\dagger)_{e e} \\
\mathcal{P}_{\mu e}^\text{int} = (N N^\dagger N N^\dagger)_{e\mu} (N N^\dagger)_{\mu e}.
\end{eqnarray}
Note that, in general, the slope of the differential cross-section in the presence of non-unitarity differs from that expected in the SM, as it changes the relative weight between the  $T$-dependent and constant terms. This is determined by the actual values of the probability factors, in turn specified by the non-unitarity parameters $\alpha_{ij}$. 
While the exact expression in Eq.~\ref{eq:diffcrossnumuNU} is a complicated function of the $\alpha_{ij}$, one can write it in powers of the seesaw expansion parameter $\varepsilon$ by following the prescription given in Eqs.~\ref{eq:epsorder1} and \ref{eq:epsorder2}. 
This implies that, to $\mathcal{O}\left(\varepsilon^2\right)$, the SM kinematic structure is preserved, and one has just an overall re-scaling compared to the SM expectation
\begin{eqnarray}
\label{eq:diffcrossnumuNUapprox}
    \left(\frac{d\sigma}{dT}\right)^\text{NU} &\approx&  \left(2\alpha_{22}^2 - \alpha_{11}^2\right) \left(\frac{d\sigma}{dT}\right)^\text{SM} + \mathcal{O}\left(\varepsilon^4\right) .
\end{eqnarray}
This can be understood by expanding the NC probability factor in powers of $\varepsilon$ and noticing that the leading term is of order $1-\mathcal{O}\left(\varepsilon^2\right)$ and can be written as 
\begin{equation}
    \mathcal{P}_{\mu e}^{\text{NC}} =  (N N^\dagger N N^\dagger N N^\dagger)_{\mu \mu}\approx 3 \alpha_{22}^2 - 2 \sim 1- \mathcal{O}\left(\varepsilon^2\right)~, \\
\end{equation}
while the CC and interference probability factors are instead $\mathcal{O}\left(\varepsilon^4\right)$, i.e.
\begin{eqnarray}
    \mathcal{P}_{\mu e}^{\text{CC}} = (N N^\dagger)_{\mu e} (N N^\dagger)_{e \mu} (N N^\dagger)_{ee} = \alpha_{11}^4 |\alpha_{21}|^2 \sim \mathcal{O}\left(\varepsilon^4\right) \\
    \mathcal{P}_{\mu e}^\text{int} = (N N^\dagger N N^\dagger)_{e\mu} (N N^\dagger)_{\mu e} \approx 2|\alpha_{21}|^2 \sim  \mathcal{O}\left(\varepsilon^4\right),
\end{eqnarray}

Therefore, up to $\mathcal{O}\left(\varepsilon^2\right)$ terms one can neglect the CC and interference contributions, hence recovering the same kinematic structure characteristic of the SM and given by Eq.~\ref{eq:diffcrossnumu}. 
Deviations from the SM kinematic structure come only at order $\mathcal{O}\left(\varepsilon^4\right)$, due to the interplay of both diagrams in Fig.~\ref{fig:numudiags}, involving genuine flavour violating NU parameters $\alpha_{ij}, i\neq j$.\par
All in all, the ratio between the expected number of events in the unitary ($\mathcal{N}_U$) and the non-unitary ($\mathcal{N}_\text{NU}$) cases is given by 
\begin{equation}
\label{eq:Neventsneutraloscillations}
    \colorbox{lightblue}{$\displaystyle\frac{\mathcal{N}_\text{NU}}{\mathcal{N}_\text{U}} = \mathcal{P}_{\mu e}^{\text{NC}} \frac{G_F^2}{G_\mu^2} + \mathcal{O}\left(\varepsilon^4\right) = \frac{(N N^\dagger N N^\dagger N N^\dagger)_{\mu \mu}}{ (N N^\dagger)_{ee} (N N^\dagger)_{\mu \mu}} \approx 2 \alpha_{22}^2- \alpha_{11}^2$} .
\end{equation}
Notice that, contrary to naive expectations, Eq.~\ref{eq:Neventsneutraloscillations} can be either bigger or smaller than $1$,  
due to the effect of the redefinition of $G_F$. Ideally, though, with very high statistics, one would be sensitive to the $T$-dependence change in the differential cross-section at $\mathcal{O}\left(\varepsilon^4\right)$ of Eq.~\ref{eq:diffcrossnumuNU}. \\

\black
\subsection{Neutrino-induced muon production} 
For sufficiently energetic ($E_\nu > 10$ GeV) incoming neutrinos, an additional process with a muon in the final state 
becomes kinematically possible. 
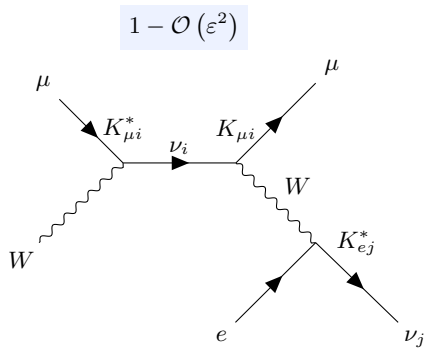
\begin{figure}[h!]
\begin{tikzpicture}
 \begin{feynman}
    \vertex (a) {\(\mu\)};
    \vertex [below right=of a] (b) ;
    \vertex [right=of b] (c) ;
    \vertex [above right=of c] (d) {\(\mu\)};
    \vertex [below left=of b] (e) {\(W\)};
    \vertex [below right=of c] (f);
    \vertex [below left=of f] (g) {\(e\)};
    \vertex [below right=of f] (h) {\(\nu_j\)};
    \diagram* {
      (a) -- [fermion] (b) -- [fermion, edge label=\(\nu_i\)] (c) -- [fermion] (d),
      (e) -- [boson] (b),
      (c) -- [boson, edge label=\(W\)] (f),
      (g) -- [fermion] (f) -- [fermion] (h),
    };
    \vertex [above=0.5em of b] {\(K_{\mu i}^*\)};
    \vertex [above=0.5em of c] {\(K_{\mu i}\)};
    \vertex [right=0.5em of f] {\(K^*_{ej}\)};
    \node [above=of b, yshift=0em, xshift=2.5em, fill=lightblue]{$1-\mathcal{O}\left(\varepsilon^2\right)$}; 
  \end{feynman}
  \end{tikzpicture}
\caption{Feynman diagram for the charged current process process \(\nu_\mu + e^- \to \nu + \mu^-\) in the non-unitary case.}
\label{fig:CCdiags1}
\end{figure}

\noindent
As shown in Fig.~\ref{fig:CCdiags1}, there is only one (CC) diagram contributing to this process.
Therefore, its kinematical structure is not modified at any order in the seesaw expansion, 
\begin{equation}
    \sigma^\text{(SM)} \approx  \frac{G_F^2}{\pi} \left(2 E_\nu m_e-m_\mu^2\right),
\end{equation}
so that the effect of non-unitarity translates only into an overall probability factor
\begin{equation}
    \mathcal{P}_{\mu \mu} =  (N N^\dagger)^2_{\mu \mu} (N N^\dagger)_{ee} \approx 2 \alpha_{22}^2+\alpha_{11}^2 - 2 \sim 1- \mathcal{O}\left(\varepsilon^2\right) \\
\end{equation}
Taking into account also the redefinition of $G_F$ we get 
\begin{equation}
\label{eq:Neventschargedoscillations}
    \colorbox{lightblue}{$\displaystyle\frac{\mathcal{N}_\text{NU}}{\mathcal{N}_\text{U}} = \mathcal{P}_{\mu \mu} \frac{G_F^2}{G_\mu^2} = (N N^\dagger)_{\mu \mu} \approx \alpha_{22}^2 \leq 1$}
\end{equation}

In summary, leptonic probes such as electron-neutrino elastic scattering and neutrino-induced muon production have a rich phenomenology and a theoretically interesting structure that allows us 
to probe non-unitarity effects in a unique manner, as shown in Eqs.~\ref{eq:diffcrossnumuNU}, \ref{eq:Neventsneutraloscillations} and \ref{eq:Neventschargedoscillations}. \\

\section{testing non-unitarity in the neutral current at a near detector} 
\label{sec:NU-NCtest}
Long-Baseline neutrino experiments require a near detector for a reliable flux calibration. They may give the opportunity to measure neutrino-electron scattering with good statistics. 
For example, DUNE is a planned particle physics experiment aimed at conducting in-depth studies of neutrinos. 
Scheduled to be hosted by Fermilab in the United States, the project's design includes sending a high-intensity neutrino beam over a distance of approximately 1,300 km from Fermilab in Illinois to a massive liquid-argon time-projection chamber in South Dakota. 
This far detector will be situated 1.5 km underground at the Sanford Underground Research Facility and is expected to have a total mass of around 40,000 metric tons. 
DUNE's main goals include investigating neutrino oscillations, exploring CP violation in the leptonic sector, and conducting astrophysical neutrino studies. 

Several near detectors are under consideration for DUNE. We will focus our analysis on a detector with similar characteristics as that of the Liquid Argon Near-Detector (ND-LAr) located at a distance from the source similar to the one considered at DUNE. 
In general, near detectors are thought to measure the neutrino beam with high precision. 
This is essential for reducing systematic uncertainties in the data to be collected by any LBL far detector. \par
The two detectors will work in tandem, with the near-far configuration enabling more accurate and reliable data interpretation. By providing initial measurements and helping calibrate the far detector's data, DUNE near detectors will be integral to achieving the scientific aims of the DUNE experiment.

The neutrino beam at Fermilab, serving as the neutrino source, can operate in two distinct modes: neutrino and antineutrino. In each mode, the generated beam contains a contribution of four neutrino types: electron neutrinos, electron antineutrinos, muon neutrinos and muon antineutrinos. 
Depending on the operation mode, the main contribution is either muon neutrinos or muon antineutrinos in the beam. The main component of the flux, the muon (anti)neutrino, peaks at around 3 GeV, but also contains a long tail extending from 10 GeV to approximately 50 GeV. \par 
In this article, we will delve into a near detector sensitivity to non-unitarity parameters through leptonic processes, taking into account specific design features mentioned above.
In order to do so we will analyze the number of events in two distinct processes that can be measured separately: neutrino-electron elastic scattering and neutrino-induced muon production. 
Notice that the flavour composition of the incoming neutrino is given by the standard flux calculation, while the outgoing neutrinos cannot be distinguished individually and thus we sum over all the kinematically accessible mass states. Moreover, we will restrict ourselves to the leading effects in the seesaw expansion. For example, probing the terms in Eq.~\ref{eq:diffcrossnumuNU} that modify the SM kinematical dependence would require measuring the recoil energy of the electrons with a precision way beyond the current envisaged experimental sensitivities for the proposed detectors. As a result, within this approximation, all the effects will depend only on the flavour conserving diagonal couplings $\alpha_{ii}$, and not on the flavour violating ones ($\alpha_{ij}$, $i\neq j $), as argued in Sec.~\ref{sec:scatt} and App.~\ref{ap:neutrinogeneral}, see for example Eq.~\ref{eq:diffcrossnumuNUapprox}.  \\ 
\subsection{Elastic neutrino-electron scattering}
Although the (anti)muon flavour dominates the flux in the (anti)neutrino mode, there are non-zero components of the 4 flavours/antiflavours involving $e$ and $\mu$. 
For each process, the number of events in the presence of lepton non-unitarity is calculated as 
\begin{equation}
    N^e_a = \mathcal{P}_a \, \mathcal{E} \int_{T_\text{min}}^{T_\text{max}(E_\nu)} \int_{E_\nu^\text{min}(T')}^\infty \frac{d\sigma_a}{dT'}  (E_\nu', T') \lambda_a(E_\nu) dE_\nu' dT',
\end{equation}
where the index $a$ runs over $\{e, \bar{e}, \mu, \bar{\mu}\}$ and $T$ is the detected electron kinetic energy, with the energy threshold in the detector being $T_\text{min} = 0.2$ GeV. 
Here $\lambda_a(E_\nu)$ is the neutrino flux, while $\mathcal{P}_a$ denotes the probability factor of that particular flavour. 
After neglecting the electron mass compared to $T$ and $E_\nu$ we obtain $E_\nu^\text{min}(T) = \frac{1}{2} (T_\text{min} + \sqrt{T_\text{min}^2 + 2 T_\text{min} m_e)} \approx T_\text{min} = 0.2 \text{ GeV}$ and $T_\text{max}(E_\nu) = \frac{2 E_\nu^2}{2 E_\nu + m_e}\approx E_\nu$. 
Finally, $\mathcal{E}$ is the exposure, calculated as the product of the number of protons on target per year $1.1 \times 10^{21} \text{ POT/year}$, the number of target electrons $N_t = 2 \times 10^{31}$ and the time spent in each mode $t= 3.5 \text{ years}$. 
The resulting numbers are given in Tab.~\ref{tab:resultse}. We present the expected number of events coming from each neutrino species within the unitary case as well as in the presence of non-unitarity.
The probability factor includes the direct effect of the non-unitary lepton mixing matrix as well as the redefinition of $G_F$, expanded to the first order in the seesaw expansion parameter $\varepsilon$. 
\begin{table}[h!]
\centering
\begin{tabular}{c c c c c c c c}
\rowcolor{lightblue} 
$\mathcal{N}_\text{U}$ & \multicolumn{2}{c}{$\nu$ mode} & \multicolumn{2}{c}{$\bar{\nu}$ mode} & $\mathcal{N}_\text{NU} / \mathcal{N}_\text{U}$ & Seesaw order & Main contribution \\
\rowcolor{lightblue}
 $\nu_a + e^- \to \nu_j + e^-$  & \textbf{events} & $\boldsymbol{\sigma}$ & \textbf{events} & $\boldsymbol{\sigma}$ &  $\mathcal{P} \, G_F^2/G_\mu^2$ & & \\
\hline \hline
\rowcolor{lightgray}
  $\nu_e$ & $2.800$ & $80$ & $1.530$ & $50$ & $\quad 2 \alpha_{11}^2-\alpha_{22}^2 \quad$ & $1\pm \mathcal{O}\left(\varepsilon^2\right)$ & NC + CC \\
\rowcolor{lightgray}
  $\nu_\mu$ & $31.400$ & $700$ & $5.800$ & $100$ & $2 \alpha_{22}^2-\alpha_{11}^2$  & $1\pm \mathcal{O}\left(\varepsilon^2\right)$ & NC\\
\rowcolor{white}
  $\bar{\nu}_{e}$  & $430$ & $20$ & $780$ & $30$  & $2 \alpha_{11}^2-\alpha_{22}^2$ & $1\pm \mathcal{O}\left(\varepsilon^2\right)$ &  NC + CC\\
\rowcolor{white}
  $\bar{\nu}_\mu$  & $3.200$ & $80$ & $20.000$ & $400$ & $2 \alpha_{22}^2-\alpha_{11}^2$  & $1\pm \mathcal{O}\left(\varepsilon^2\right)$ & NC \\
\rowcolor{lightgray}
  \textbf{total}  & $37.800$ & $800$ & $28.000$ & $600$
\end{tabular}
\caption{Expected number of electron events in the unitary case coming from each relevant neutrino type. The detector is sensitive only to the total number of events, but the probability factor in the presence of non-unitarity is different for each incoming (anti)flavour. The error $\sigma$ is taken as $\sigma^2 = \sigma_\text{stat}^2 + \sigma_\text{syst}^2$, where the statistical uncertainty $\sigma_\text{stat} = \sqrt{N}$ and the systematic uncertainty is taken as $2\%$, see text for details.}
\label{tab:resultse}
\end{table}

\subsection{ Neutrino-induced muon production} 

In order to be kinematically allowed, muon production requires more energetic neutrinos,
with the kinematic threshold given by $E_\nu > (m_\mu^2-m_e^2) / 2 m_e \approx 10$ GeV. 
While the flux peaks at a lower energy $\sim 3 \text{ GeV}$, the tails can still generate a significant number of muon events in the final state, thus improving the sensitivity to the non-unitarity parameters. The number of events is given by
\begin{equation}
    N^\mu_a = \mathcal{P}_a \, \mathcal{E} \int_{E_\nu^\text{min}}^\infty \sigma(E_\nu) \lambda_a(E_\nu) dE_\nu~,
\end{equation}
where, again, $\mathcal{E}$ is the exposure and $\lambda_a(E_\nu)$ is the flux. 
\begin{table}[h!]
\centering
\begin{tabular}{c c c c c c c c}
\rowcolor{lightblue} 
$\mathcal{N}_\text{U}$ & \multicolumn{2}{c}{$\nu$ mode} & \multicolumn{2}{c}{$\bar{\nu}$ mode} & $\mathcal{N}_\text{NU} / \mathcal{N}_\text{U}$ & Seesaw order & Main contribution \\
\rowcolor{lightblue}
 $\nu_a + e^- \to \nu_j+ \mu^-$ & \textbf{events} & $\boldsymbol{\sigma}$ & \textbf{events} & $\boldsymbol{\sigma}$ & $\mathcal{P} \, G^2_F/G^2_\mu$ & & \\
\hline \hline
\rowcolor{lightgray}
  $\nu_e$ & $0$ & $0$ & $0$ & $0$ & $|\alpha_{21}|^2$ & $ \mathcal{O}\left(\varepsilon^4\right)$ & $ \mathcal{O}\left(\varepsilon^4\right)$\\
\rowcolor{lightgray}
  $\nu_\mu$ & $17.900$ & $400$ & $14.200$ & $300$ & $\alpha_{22}^2$& $1- \mathcal{O}\left(\varepsilon^2\right)$ & CC\\
\rowcolor{white}
  $\bar{\nu}_{e}$  & $380$ & $20$ & $230$ & $20$ &$\alpha_{11}^2$ & $1- \mathcal{O}\left(\varepsilon^2\right)$ & CC\\
\rowcolor{white}
  $\bar{\nu}_\mu$  & $0$ & $0$ & $0$ & $0$ & $|\alpha_{21}|^2$ & $ \mathcal{O}\left(\varepsilon^4\right)$ &  $ \mathcal{O}\left(\varepsilon^4\right)$\\
\rowcolor{lightgray}
  \textbf{total}  & $18.300$ & $400$ & $14.400$ & $300$ 
\end{tabular}
\caption{Expected number of muon events in the unitary case coming from each species. The detector is sensitive only to the total event number, but the probability factor in the presence of non-unitarity differs for each incoming (anti)flavour. $\sigma$ is $\sigma^2 = \sigma_\text{stat}^2 + \sigma_\text{syst}^2$, where $\sigma_\text{stat} = \sqrt{N}$ and the systematic uncertainty is taken as $2\%$, see text for details.}
\label{tab:resultsmu}
\end{table}

Note that now $a$ denotes either the incoming muon neutrino or electron antineutrino. Muon antineutrino and electron neutrino flavours do not contribute to this process in the standard unitary case while, in the presence of non-unitarity, their contribution appears only as order $\mathcal{O}\left(\varepsilon^4\right)$ corrections. The results are given in Tab.~\ref{tab:resultsmu}. 

%
\begin{figure}[t!]
\begin{adjustbox}{center}
\includegraphics[height=7.5cm]{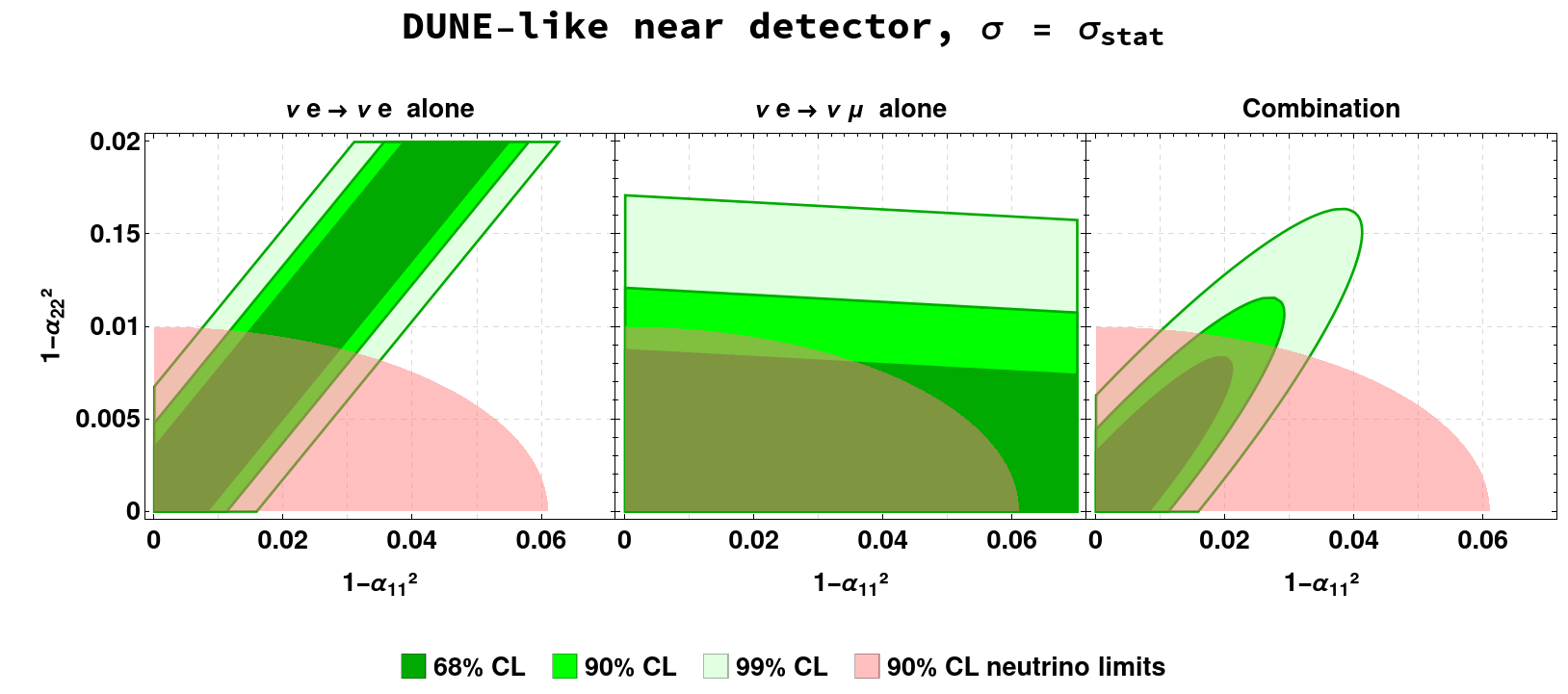}
\end{adjustbox}
                \caption{Sensitivities on the parameters $\alpha_{11}$ and $\alpha_{22}$ from our suggested \jv{DUNE}-like near detector. The neutral current-dominated elastic neutrino-electron scattering process constrains the quantity $2\alpha_{22}^2 - \alpha_{11}^2$ (left panel), while the purely charged current process $\nu_\mu + e^- \to \nu_j + \mu^-$ involves only $\alpha_{22}^2$ (middle panel). The combination of both event types in the lepton channel constrains both non-unitarity parameters $\alpha_{11}^2$ and $\alpha_{22}^2$ (right pannel). As a benchmark, we show the sensitivity of a hypothetical experiment where the statistical uncertainty dominates over the systematics. For comparison, the pink region represents the constraints from both long and short baseline experiments, as detailed in \cite{Forero:2021azc}.}
\label{fig:results1} 
\end{figure}
\subsection{Analysis}
\begin{figure}[t!]
\begin{adjustbox}{center}
\includegraphics[height=7cm]{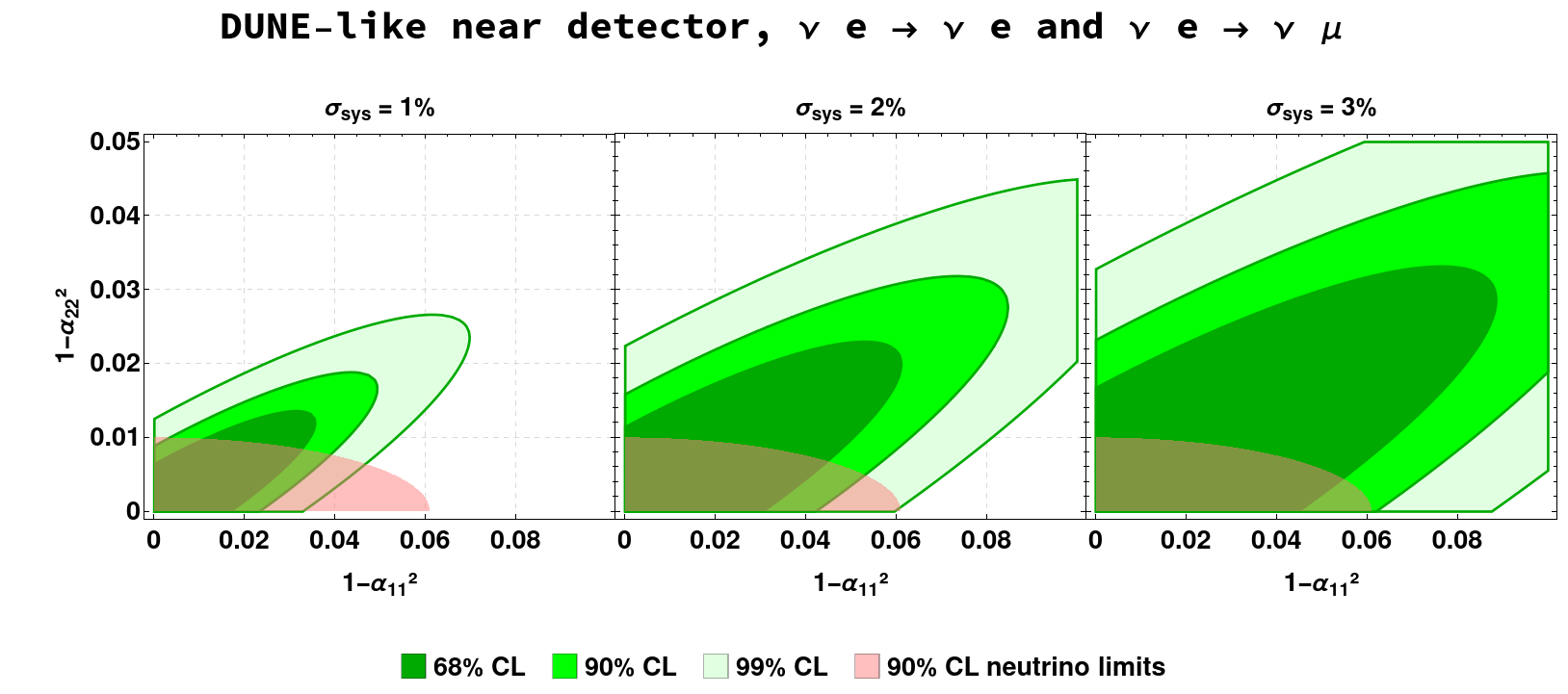}
\end{adjustbox} 
       \caption{Combination of the NC ($\nu + e \rightarrow \nu + e$) and CC ($\nu + e \rightarrow \nu + \mu$) results for different values of the systematic uncertainty. Again, the pink region represents the constraints from both long and short baseline experiments \cite{Forero:2021azc}. While the constraints on $\alpha_{22}^2$ are already strong, this setup has the potential to significantly improve also the one on $\alpha_{11}^2$. }
\label{fig:NearDune123}
\end{figure}
We now proceed to estimate the sensitivity of our proposed leptonic-probe experiment to the non-unitarity parameters. In order to do so we take the $\chi^2$ function as 
\begin{equation}
    \chi^2 = \sum_i\frac{(N^\text{the}_i - N_i^\text{exp})^2}{\sigma^2_{i, \text{ stat}}+\sigma^2_{i, \text{ syst}}},
\end{equation}
where the subscript $i$ runs over the $4$ possible measurements: electron or muon events during the neutrino or antineutrino modes. 
Here $N_i^\text{exp}$ is the number of detected events, while $N_i^\text{the}$ is the expected number of events given as a function of the non-unitarity parameters and $\sigma^2_{i, \text{ stat}} = N_i^\text{exp}$. \par
In the analysis we will take different possible values for the systematic error, ranging from $0-3\%$, in order to account for the uncertainty in the flux calculation. 
We are aware that our benchmark assumption is too optimistic for the current setup, but it should serve as motivation for improvement, given the interest of the associated physics.   

Notice that non-unitarity accounts for the fact that the heavy mediator states are not kinematically accessible.
The direct effects of non-unitarity tend to decrease the event number, and this goes
in the opposite direction as the redefinition of the Fermi constant $G_F$ discussed above.
As a result, there is an intrinsic ambiguity in the electron events alone, leading to a parameter degeneracy. 
Note however that this ambiguity is not present in the charged current events, which are only sensitive to $\alpha^2_{22}$. 
Hence one can obtain a limit for both $\alpha_{11}^2$ and $\alpha_{22}^2$ only combining CC and NC event types. 
This behaviour can be seen in Fig.~\ref{fig:results1} where, for illustration purposes, an optimistic $\sigma = \sigma_\text{stat}$ uncertainty was assumed. 
We compare the recently updated ``neutrino limits'' extracted from neutrino oscillations at both long and short baseline experiments~\cite{Hu:2020oba,Ellis:2020hus,Forero:2021azc,Agarwalla:2021owd}. \\

A few additional comments are in order. A realistic sensitivity estimate would require a detailed study of the background. In several background analysis discussions of neutrino-electron scattering experiments, a common cut is to select events with small values for the $E\theta^2$ variable (see for instance~\cite{MINERvA:2022vmb}).
%
\par On the other hand, hadronic processes will also be sensitive to non-unitarity effects~\cite{Miranda:2018yym} and will dominate the statistics compared with neutrino electron scattering (for a recent discussion see~\cite{DUNE:2021tad}). The results of both analyses will be complementary, as the dependence on the non-unitarity parameters will be different. In particular, the effects discussed here are $K^6$ and involve an interference between the CC and NC, while the hadronic processes are either purely NC and $K^6$, or purely CC and $K^4$. Note that, despite having smaller statistics, neutrino electron scattering will be a cleaner process as it is purely weak, free of strong corrections, so its cross-section uncertainties will be smaller than for the hadronic case. 

In any case, our goal here is to focus more on the novelty of the measurement, rather that on the strength of the resulting sensitivities, as the setup is currently not optimized to do so. 
However, we can see that the attainable results can already be comparable to other similar experiments. In particular, the $\alpha_{11}^2$ can be probed significantly better than current oscillation experiments~\cite{Forero:2021azc}. 
It is worth noting that in general there are constraints arising from universality tests \cite{Bryman:2021teu}, electroweak precision measurements \cite{Workman:2022ynf}, such as the invisible Z decay \cite{Escrihuela:2019mot}, and charged Lepton Flavor Violation searches~\cite{Forero:2011pc,Lindner:2016bgg}.
For a recent extensive discussion see \cite{Blennow:2023mqx}.
While the combination of these restrictions can be rather stringent, we stress the cleanliness of our proposed leptonic probe, and the fact that the interpretation is fairly model-independent. 

In Fig.~\ref{fig:NearDune123} we now consider the role of different values of the systematic uncertainty and again compare the resulting sensitivities with the current oscillation constraint.

\begin{figure}[h!]
\includegraphics[height=10cm]{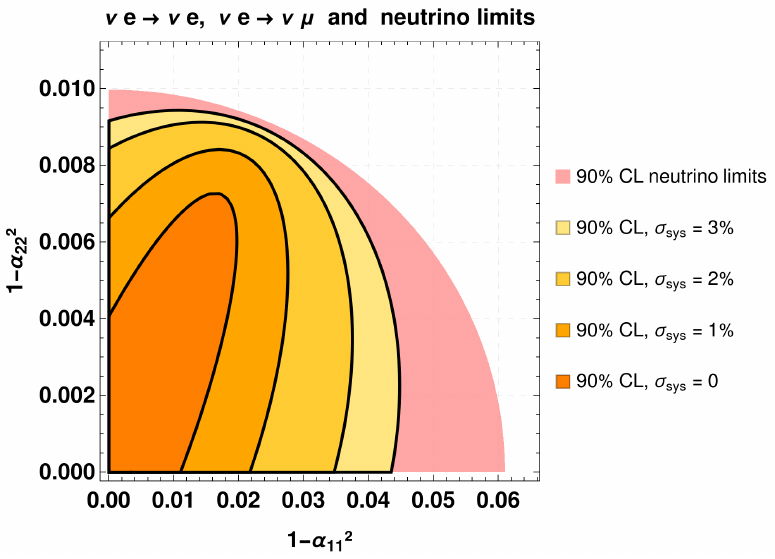} 
       \caption{Combination of our proposed DUNE-like near detector measurements, i.e. neutrino electron scattering and neutrino induced muon production, with the long and short baseline oscillation experiments constraints of \cite{Forero:2021azc} at 90\% CL for different values of the systematic uncertainty.}
\label{fig:combination}
\end{figure}

Finally, we note that by combining the results from a DUNE-like near-detector with the oscillation constraints for different values of the systematic uncertainty, the measurement for $\alpha_{11}^2$ can indeed improve, see Fig.~\ref{fig:combination}.
The fact that this happens even when the setup is not optimized for this type of measurement should serve as a motivation for the design of future experiments using leptonic neutral current as a way to underpin the neutrino mass generation seesaw mechanism. 

\section{Conclusions and outlook}

Here we have proposed the use of leptonic probes, such as neutrino-electron scattering, as a viable way to test the non-minimal form of the charged and neutral leptonic weak interactions within a DUNE-like near-detector setup.
Although the statistics is low, these processes are very clean and can provide complementary information to that available from oscillation studies. 
While the current setup is not optimized to our proposal, our results already indicate that there is potential for significant improvement on the sensitivities for the non-unitarity parameter $\alpha_{11}^2$ when compared with current oscillation experiments. 
For example, Figs.~\ref{fig:results1},~\ref{fig:NearDune123} and \ref{fig:combination} illustrate how our method can help improving non-unitarity constraints and thereby shed light on the scale of neutrino mass generation within low-scale seesaw schemes.
Last, but not least, our results further highlight the fact that a robust experimental setup for neutrino research requires the presence of near detectors. In addition to ensuring robustness, these short-distance studies can provide, by themselves, valuable information on new physics parameters, such as the nonunitarity parameters, which constitutes an interesting physics goal by itself.
Besides a plethora of low-energy probes~\cite{Miranda:2021kre,Schwetz:2021thj,Schwetz:2021cuj,Tang:2021lyn,Arrington:2022pon,Capozzi:2023ltl,Soleti:2023hlr}, there are plenty of physics opportunities for testing unitarity violation. For example, through the searches for charged lepton flavor violating processes at low and high-energies
\cite{Bernabeu:1987gr,Gonzalez-Garcia:1988okv,Abada:2014kba,Hagedorn:2021ldq}.
One may also have the possibility of directly producing the TeV-scale neutrino-mass-mediators at colliders~\cite{Dittmar:1989yg, Gonzalez-Garcia:1990sbd, Atre:2009rg,Aguilar-Saavedra:2012dga,Das:2012ii,Deppisch:2013cya,Antusch:2015mia,Deppisch:2015qwa,Hirsch:2020klk,Cottin:2022nwp,Chauhan:2023faf,Batra:2023mds}.
The rich variety of associated signals justifies the intense experimental effort devoted in present and upcoming experiments~\cite{Alekhin:2015byh,CMS:2018szz,ATLAS:2018dcj,Alimena:2019zri,CMS:2022fut,Abdullahi:2022jlv,CMS:2023jqi}.

 \black

\begin{acknowledgments}
  We thank Joachim Kopp and Jaehoon Yu for insightful discussions. This work was supported by the Spanish grants PID2020-113775GB-I00~(AEI/10.13039/501100011033) and Prometeo CIPROM/2021/054 (Generalitat Valenciana). OGM was supported by the CONAHCyT (Consejo Nacional de Humanidades ciencias y tecnologías) grant 23238 and by SNII (Sistema Nacional de Investigadoras e Investigadores)-M\'exico. 
\end{acknowledgments}

\appendix

\section{General formalism for neutrino and antineutrino scattering }
\label{ap:neutrinogeneral}

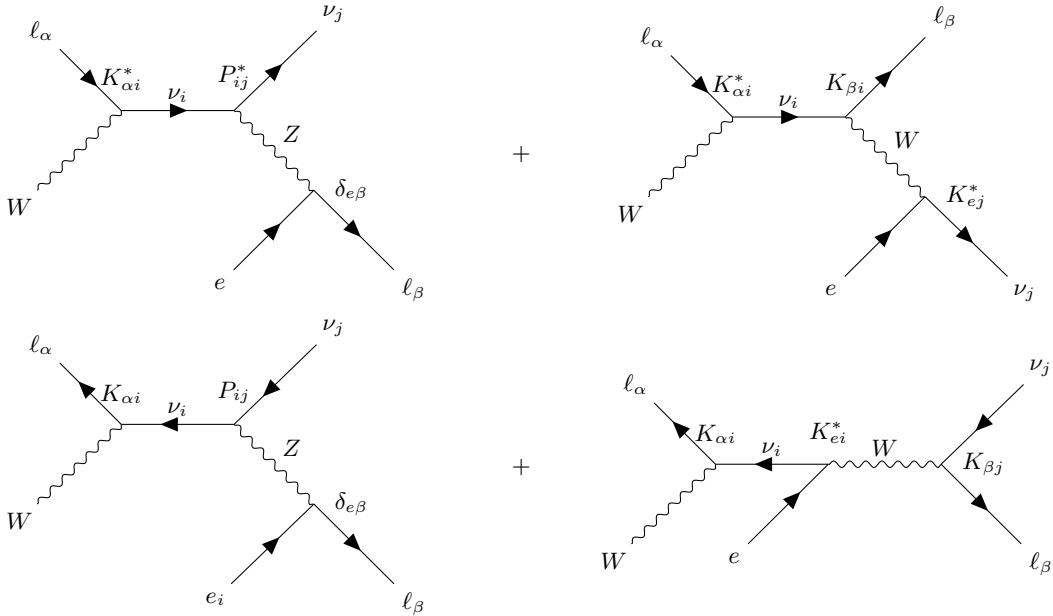
\begin{figure}[t!]
  \centering
  \begin{minipage}{0.45\linewidth}
    \centering
    \begin{tikzpicture}
      \begin{feynman}
\vertex (a) {\(\ell_\alpha\)};
    \vertex [below right=of a] (b) ;
    \vertex [right=of b] (c) ;
    \vertex [above right=of c] (d) {\(\nu_j\)};
    \vertex [below left=of b] (e) {\(W\)};
    \vertex [below right=of c] (f);
    \vertex [below left=of f] (g) {\(e\)};
    \vertex [below right=of f] (h) {\(\ell_\beta\)};
    \diagram* {
      (a) -- [fermion] (b) -- [fermion, edge label=\(\nu_i\)] (c) -- [fermion] (d),
      (e) -- [boson] (b),
      (c) -- [boson, edge label=\(Z\)] (f),
      (g) -- [fermion] (f) -- [fermion] (h),
    };
    \vertex [above=0.5em of b] {\(K^*_{\alpha i}\)};
    \vertex [above=0.5em of c] {\(P^*_{ij}\)};
    \vertex [right=0.5em of f] {\(\delta_{e\beta}\)};
      \end{feynman}
    \end{tikzpicture}
  \end{minipage}
  \(+\)
  \begin{minipage}{0.45\linewidth}
    \centering
    \begin{tikzpicture}
       \begin{feynman}
  \vertex (a) {\(\ell_\alpha\)};
    \vertex [below right=of a] (b) ;
    \vertex [right=of b] (c) ;
    \vertex [above right=of c] (d) {\(\ell_\beta\)};
    \vertex [below left=of b] (e) {\(W\)};
    \vertex [below right=of c] (f);
    \vertex [below left=of f] (g) {\(e\)};
    \vertex [below right=of f] (h) {\(\nu_j\)};
    \diagram* {
      (a) -- [fermion] (b) -- [fermion, edge label=\(\nu_i\)] (c) -- [fermion] (d),
      (e) -- [boson] (b),
      (c) -- [boson, edge label=\(W\)] (f),
      (g) -- [fermion] (f) -- [fermion] (h),
    };
    \vertex [above=0.5em of b] {\(K^*_{\alpha i}\)};
    \vertex [above=0.5em of c] {\(K_{\beta i}\)};
    \vertex [right=0.5em of f] {\(K^*_{ej}\)};
      \end{feynman} 
    \end{tikzpicture}
  \end{minipage}

  \begin{minipage}{0.45\linewidth}
    \centering
    \begin{tikzpicture}
      \begin{feynman}
    \vertex (a) {\(\ell_\alpha\)};
    \vertex [below right=of a] (b) ;
    \vertex [right=of b] (c) ;
    \vertex [above right=of c] (d) {\(\nu_j\)};
    \vertex [below left=of b] (e) {\(W\)};
    \vertex [below right=of c] (f);
    \vertex [below left=of f] (g) {\(e_i\)};
    \vertex [below right=of f] (h) {\(\ell_\beta\)};
    \diagram* {
      (a) -- [anti fermion] (b) -- [anti fermion, edge label=\(\nu_i\)] (c) -- [anti fermion] (d),
      (e) -- [boson] (b),
      (c) -- [boson, edge label=\(Z\)] (f),
      (g) -- [fermion] (f) -- [fermion] (h),
    };
    \vertex [above=0.5em of b] {\(K_{\alpha i}\)};
    \vertex [above=0.5em of c] {\(P_{ij}\)};
    \vertex [right=0.5em of f] {\(\delta_{e\beta}\)};
  \end{feynman} 
    \end{tikzpicture}
  \end{minipage}
  \(+\)
  \begin{minipage}{0.45\linewidth}
    \centering
    \begin{tikzpicture}
   \begin{feynman}
    \vertex (a) {\(\ell_\alpha\)};
    \vertex [below right=of a] (b) ;
    \vertex [right=of b] (c) ;
    \vertex [below left=of c] (d) {\(e\)};
    \vertex [below left=of b] (e) {\(W\)};
    \vertex [right=of c] (f);
    \vertex [above right=of f] (g) {\(\nu_j\)};
    \vertex [below right=of f] (h) {\(\ell_\beta\)};
    \diagram* {
      (a) -- [anti fermion] (b) -- [anti fermion, edge label=\(\nu_i\)] (c) -- [anti fermion] (d),
      (e) -- [boson] (b),
      (c) -- [boson, edge label=\(W\)] (f),
      (g) -- [fermion] (f) -- [fermion] (h),
    };
    \vertex [above=0.5em of b] {\({K_{\alpha i}}\)};
    \vertex [above=0.5em of c] {\({K^*_{ei}}\)};
    \vertex [right=0.5em of f] {\({K_{\beta j}}\)};
  \end{feynman} \hspace{8cm}
    \end{tikzpicture}
  \end{minipage}
  \caption{Interplay between charged and neutral currents in the processes $\nu_\alpha + e^- \to \nu + \ell^{-}_\beta$ (upper row) and $\bar{\nu}_\alpha + e^- \to \bar{\nu} + \ell^{-}_\beta$ (lower row). The standard probability factors for each diagram are either $1$ or $0$, depending on the initial flavour $\alpha$ and the final charged lepton $\ell_\beta$. However, in the presence of non-unitarity both diagrams are present, each carrying a different probability factor. Within a consistent seesaw expansion we show that the SM kinematic structure is preserved at order $\mathcal{O}\left(\varepsilon^2\right)$, see text for details.}  
  \label{fig:diagsgeneral}
\end{figure}

Here we consider the general family of neutrino scattering processes
$\nu_\alpha + e^- \to \nu + \ell^{-}_\beta$ and $\bar{\nu}_\alpha + e^- \to \bar{\nu} + \ell^{-}_\beta$ 
where $\alpha$ denotes the initial (anti)neutrino flavour, and the final lepton $\ell^-_\beta$ can be an electron, a muon or a tau, if kinematically possible. 
In other words, the initial (anti)neutrino is produced in association with a charged lepton of definite flavour. Moreover the neutrinos mix with new heavy mediator states.
Since the final (anti)neutrino state is not measured,we sum over the three kinematically accessible neutral mass states $j$.
We discuss the effect of non-unitarity at zero distance.  \par

At tree level there is always a non-zero contribution coming from the charged current irrespective of $\alpha$ and $\beta$. 
However, we show that the cases in which this contribution vanishes in the unitary case are actually of order $\mathcal{O}\left(\varepsilon^4\right)$ in the presence of non-unitarity.
On the other hand, when the probability factor is $1$ in the standard case one now has a modification by a common factor of order $1-\mathcal{O}\left(\varepsilon^2\right)$. 
For the NC there is only a non-zero contribution if $\beta = e$, in which case it is always of order $1-\mathcal{O}\left(\varepsilon^2\right)$.

\subsection{Neutrino scattering} 

The full structure of the relevant diagrams is shown in the upper row of Fig.~\ref{fig:diagsgeneral}, while the relevant probability (and interference) factors are given in Eqs.~\ref{eq:probfact1}-\ref{eq:probfact3}. Note that the probability factors are different for each combination of diagrams, i.e. 
\begin{eqnarray}
    \mathcal{P}_{\alpha \beta}^{\text{NC}} &=& \delta_{e \beta} (N N^\dagger N N^\dagger N N^\dagger)_{\alpha \alpha}\label{eq:probfact1} \\
    \mathcal{P}_{\alpha \beta}^{\text{CC}} &=& (N N^\dagger)_{\alpha \beta} (N N^\dagger)_{\beta \alpha} (N N^\dagger)_{ee} \label{eq:probfact2}\\
    \mathcal{P}_{\alpha \beta}^\text{int} &=& \delta_{e \beta} (N N^\dagger)_{\beta \alpha} (N N^\dagger N N^\dagger)_{\alpha e} \label{eq:probfact3}
\end{eqnarray}
Notice that $\alpha$ and $\beta$ denote the fixed flavor indices associated to the survival or conversion probability, while the latin indices $i, j$ are summed over, that is \[(N N^\dagger)_{\alpha \beta} = \sum_{j=1}^3 N_{\alpha j} N^\dagger_{j \beta}.\] 
Each of the probability factors in Eqs.~\ref{eq:probfact1}-\ref{eq:probfact3} affects different terms in the cross section, making the complete expression a complicated one. However, as already discussed for the particular $\nu_\mu$-electron elastic scattering case in Sec.~\ref{subsec3a}, one finds that in general the kinematic structure of the SM is preserved at order $\mathcal{O}\left(\varepsilon^2\right)$. Modifications to the SM spectrum shape only come at order $\mathcal{O}\left(\varepsilon^4\right)$. For example, up to $\mathcal{O}\left(\varepsilon^2\right)$, one finds
\begin{align}
    \mathcal{P}_{\alpha \beta}^{\text{NC}} &\approx \delta_{e \beta} (3 \alpha^2_{\alpha \alpha} - 2) + \mathcal{O}\left(\varepsilon^4\right) \sim \delta_{e \beta} \times \left[1-\mathcal{O}\left(\varepsilon^2\right)\right] \label{eq:probfact11}\\
    \mathcal{P}_{\alpha \beta}^{\text{CC}} &\approx \delta_{\alpha \beta} (2\alpha^2_{\alpha \alpha}+\alpha^2_{11}-2)+ \mathcal{O}\left(\varepsilon^4\right) \sim \delta_{\alpha \beta} \times \left[1-\mathcal{O}\left(\varepsilon^2\right)\right] \label{eq:probfact22}\\
    \mathcal{P}_{\alpha \beta}^{\text{int}} &\approx \delta_{e\beta} \delta_{\alpha \beta} (3 \alpha_{11}^2 -2)+ \mathcal{O}\left(\varepsilon^4\right) \sim \delta_{e\beta} \delta_{\alpha \beta} \times \left[1-\mathcal{O}\left(\varepsilon^2\right)\right] \label{eq:probfact33}
\end{align}

\black
From these one can easily conclude:
\begin{itemize}
    \item In inelastic scattering processes (muon or tau neutrino-induced production), where the final charged lepton is not an electron, only the CC contribution exists. This corresponds to the case $\beta \neq e$.
    \item In elastic neutrino-electron scattering, $\beta=e$, if the initial neutrino flavour is of $\mu$ or $\tau$ type, $\alpha\neq e$, there is a CC contribution of order $\mathcal{O}\left(\varepsilon^4\right)$, subleading in comparison to the neutral current.
    \item If $\beta=e$ and the initial neutrino flavour is of electron type, $\alpha= e$, then the probability factors of the NC, the CC and the interference are all equal to $3\alpha_{11}^2-2$. This means that, again, there is no spectral distortion at order $\mathcal{O}\left(\varepsilon^2\right)$.
\end{itemize}

We emphasize that at tree-level and $\mathcal{O}\left(\varepsilon^2\right)$ in the seesaw expansion parameter the effect of non-unitarity is just a global factor times the SM prediction. This conclusion no longer holds at second order in the seesaw expansion, when the flavour-violating parameters of Eq.~\ref{eq:Nparam} come into play. In this case the $\mathcal{N}_\text{NU}/\mathcal{N}_\text{U}$ event number ratio will not be a constant and will instead depend on the kinematic parameters and flux shape.

A summary of these results is given in Tab.~\ref{tab:probs2}, and the SM cross sections are given by

\begin{table}[t!]
\centering
\begin{tabular}{c c c c c c}
\rowcolor{lightblue} 
Process & Contribution & $\mathcal{P}$ factor at $\mathcal{O}\left(\varepsilon^2\right)$ & $\mathcal{N}_\text{NU}/\mathcal{N}_\text{U} = \mathcal{P} \, G_F^2/G_\mu^2$ & SM cross section & Kinematic threshold (Lab) \\
\hline \hline
\rowcolor{lightgray} 
$\nu_e + e^- \to \nu_j + e^-$& NC+CC& $3\alpha_{11}^2-2$& $2\alpha_{11}^2-\alpha_{22}^2$ & Eq.~\ref{eq:kinfac1} &  \\
\rowcolor{lightgray} 
$\nu_\mu + e^- \to \nu_j + e^-$& NC&$3\alpha_{22}^2-2$& $2\alpha_{22}^2-\alpha_{11}^2$ & Eq.~\ref{eq:kinfac2} & No \\
\rowcolor{lightgray} 
$\nu_\tau + e^- \to \nu_j + e^-$& NC& $3\alpha_{33}^2-2$& $3\alpha_{33}^2-\alpha_{22}^2 - \alpha_{11}^2$& Eq.~\ref{eq:kinfac2} & \\
\rowcolor{white} 
$\nu_e + e^- \to \nu_j +  \mu^-$& & $\mathcal{O}\left(\varepsilon^4\right)$ & $\mathcal{O}\left(\varepsilon^4\right)$  & &  \\
\rowcolor{white} 
$\nu_\mu + e^- \to \nu_j +  \mu^-$& CC & $2\alpha_{22}^2+\alpha_{11}^2-2$ & $\alpha_{22}^2$ & Eq.~\ref{eq:kinfac3}& $E_\nu > 10 \text{ GeV}$\\
\rowcolor{white} 
$\nu_\tau + e^- \to \nu_j +  \mu^-$& & $\mathcal{O}\left(\varepsilon^4\right)$ & $\mathcal{O}\left(\varepsilon^4\right)$  & & \\
\rowcolor{lightgray} 
$\nu_e + e^- \to \nu_j +  \tau^-$ & & $\mathcal{O}\left(\varepsilon^4\right)$ & $\mathcal{O}\left(\varepsilon^4\right)$  & &  \\
\rowcolor{lightgray} 
$\nu_\mu + e^- \to \nu_j +  \tau^-$ & & $\mathcal{O}\left(\varepsilon^4\right)$ & $\mathcal{O}\left(\varepsilon^4\right)$   & & $E_\nu > 3 \text{ TeV}$ \\
\rowcolor{lightgray} 
$\nu_\tau + e^- \to \nu_j +  \tau^-$ & CC & $2\alpha_{33}^2+\alpha_{11}^2-2$ & $2\alpha_{33}^2-\alpha_{22}^2$ & Eq.~\ref{eq:kinfac4} & \\
\end{tabular}
\caption{Structure of the family of processes $\nu_\alpha + e^- \to \nu + \ell_\beta^{-}$ in the presence of non-unitarity, including the probability factors at order $\mathcal{O}\left(\varepsilon^2\right)$, the kinematic factors for each contribution: NC, CC or the interference.}
\label{tab:probs2}
\end{table}

\begin{align}
    {\frac{d\sigma}{dT}}^{NC+CC}(\nu + e^- \to \nu +  e^-) &=& \frac{2 G_F^2 m_e}{\pi} \left((1+g_L)^2 + g_R^2 \left(1- \frac{T}{E\nu}\right)^2 - (1+g_L) g_R  \frac{m_e T}{E_\nu^2}\right) \label{eq:kinfac1} \\
    {\frac{d\sigma}{dT}}^{NC}(\nu + e^- \to \nu +  e^-) &=& \frac{2 G_F^2 m_e}{\pi} \left(g_L^2 + g_R^2 \left(1- \frac{T}{E\nu}\right)^2 - g_L g_R  \frac{m_e T}{E_\nu^2}\right) \label{eq:kinfac2} \\
        \sigma(\nu + e^- \to \nu +  \mu^-) &=&  \frac{G_F^2}{\pi} \left(2 E_\nu m_e-m_\mu^2\right) \label{eq:kinfac3} \\
    \sigma(\nu + e^- \to \nu +  \tau^-) &=& \frac{G_F^2}{\pi} \left(2 E_\nu m_e-m_\tau^2\right) \label{eq:kinfac4}
\end{align}

\subsection{Antineutrino scattering}
\begin{table}[t!]
\centering
\begin{tabular}{c c c c c c}
\rowcolor{lightblue} 
Process & Contribution & $\mathcal{P}$ factor at $\mathcal{O}\left(\varepsilon^2\right)$ & $\mathcal{N}_\text{NU}/\mathcal{N}_\text{U} = \mathcal{P} \, G_F^2/G_\mu^2$ & SM cross section & Kinematic threshold (Lab) \\
\hline \hline
\rowcolor{lightgray} 
$\bar{\nu}_e + e^- \to \bar{\nu}_j + e^-$& NC+CC& $3\alpha_{11}^2-2$& $2\alpha_{11}^2-\alpha_{22}^2$ & Eq.~\ref{eq:kinfac1}, $(1+g_L) \leftrightarrow g_R$ &  \\
\rowcolor{lightgray} 
$\bar{\nu}_\mu + e^- \to \bar{\nu}_j + e^-$& NC&$3\alpha_{22}^2-2$& $2\alpha_{22}^2-\alpha_{11}^2$ & Eq.~\ref{eq:kinfac2}, $g_L \leftrightarrow g_R$ & No \\
\rowcolor{lightgray} 
$\bar{\nu}_\tau + e^- \to \bar{\nu}_j + e^-$& NC& $3\alpha_{33}^2-2$& $3\alpha_{33}^2-\alpha_{22}^2 - \alpha_{11}^2$& Eq.~\ref{eq:kinfac2}, $g_L \leftrightarrow g_R$& \\
\rowcolor{white} 
$\bar{\nu}_e + e^- \to \bar{\nu}_j +  \mu^-$& CC & $2\alpha_{11}^2+\alpha_{22}^2-2$ & $\alpha_{11}^2$ & Eq.~\ref{eq:kinfac3} &  \\
\rowcolor{white} 
$\bar{\nu}_\mu + e^- \to \bar{\nu}_j +  \mu^-$& & $\mathcal{O}\left(\varepsilon^4\right)$ & $\mathcal{O}\left(\varepsilon^4\right)$   & & $E_\nu > 10 \text{ GeV}$\\
\rowcolor{white} 
$\bar{\nu}_\tau + e^- \to \bar{\nu}_j +  \mu^-$& &$\mathcal{O}\left(\varepsilon^4\right)$ & $\mathcal{O}\left(\varepsilon^4\right)$  & & \\
\rowcolor{lightgray} 
$\bar{\nu}_e + e^- \to \bar{\nu}_j +  \tau^-$ & CC & $2\alpha_{11}^2+\alpha_{33}^2-2$ &  $\alpha_{11}^2-\alpha_{22}^2+\alpha_{33}^2$&Eq.~\ref{eq:kinfac4} &  \\
\rowcolor{lightgray} 
$\bar{\nu}_\mu + e^- \to \bar{\nu}_j +  \tau^-$ & &$\mathcal{O}\left(\varepsilon^4\right)$ & $\mathcal{O}\left(\varepsilon^4\right)$   & & $E_\nu > 3 \text{ TeV}$ \\
\rowcolor{lightgray} 
$\bar{\nu}_\tau + e^- \to \bar{\nu}_j +  \tau^-$ &  & $\mathcal{O}\left(\varepsilon^4\right)$ & $\mathcal{O}\left(\varepsilon^4\right)$   &  & \\
\end{tabular}
\caption{Summary of the family of processes $\bar{\nu}_\alpha + e^- \to \bar{\nu} + \ell_\beta^{-}$ in the presence of non-unitarity, including the probability factors at order $\mathcal{O}\left(\varepsilon^2\right)$, the kinematic factors and the contribution type: NC, CC or the interference.}
\label{tab:probs3}
\end{table}
 
While the results will be very similar to the neutrino case, we include them here for completeness. The diagrams are slightly different and are given by the lower row of Fig.~\ref{fig:diagsgeneral}, while their respective probability factors (and the interference) are given in Eqs.~\ref{eq:probfactanti1}-\ref{eq:probfactanti3}. 
\begin{eqnarray}
    \mathcal{P}_{\alpha \beta}^{\text{NC}} &=& \delta_{e \beta} (N N^\dagger N N^\dagger N N^\dagger)_{\alpha \alpha}\label{eq:probfactanti1} \\
    \mathcal{P}_{\alpha \beta}^{\text{CC}} &=& (N N^\dagger)_{\alpha e} (N N^\dagger)_{e \alpha} (N N^\dagger)_{\beta \beta} \label{eq:probfactanti2}\\
    \mathcal{P}_{\alpha \beta}^\text{int} &=& \delta_{e \beta} (N N^\dagger)_{e \alpha} (N N^\dagger N N^\dagger)_{\alpha \beta} \label{eq:probfactanti3}
\end{eqnarray}
i.e. just replacing $e \leftrightarrow \beta$ in the neutrino case. Similar conclusions apply and a summary can be found in Tab.~\ref{tab:probs3}
\black
\bibliographystyle{utphys}
\providecommand{\href}[2]{#2}\begingroup\raggedright\endgroup

\end{document}